\documentclass[12pt,a4paper]{article}

\usepackage{graphicx}
\usepackage{amsmath}
\usepackage{amssymb}
\usepackage{enumerate}
\usepackage{natbib}

\newcommand{\eref}[1]{(\ref{#1})}

\newcommand{\dd}{\textrm{d}}
\newcommand{\doo}{\textrm{do}}
\newcommand{\Jyvaskyla}{Jyv{\"a}skyl{\"a}}

\newcommand\numberthis{\addtocounter{equation}{1}\tag{\theequation}}
\newtheorem{definition}{Definition}

\newcommand\independent{\protect\mathpalette{\protect\independenT}{\perp}}
\def\independenT#1#2{\mathrel{\rlap{$#1#2$}\mkern2mu{#1#2}}}

\begin{document}
\title{Study design in causal models}
\author{Juha Karvanen\\
Department of Mathematics and Statistics,\\
University of \Jyvaskyla,\\
\Jyvaskyla, Finland\\
juha.t.karvanen@jyu.fi}

\maketitle 


\section*{Abstract}
The causal assumptions, the study design and the data are the elements required for scientific inference in empirical research. The research is adequately communicated only if all of these elements and their relations are described precisely. Causal models with design describe the study design and the missing data mechanism together with the causal structure and allow the direct application of causal calculus in the estimation of the causal effects. The flow of the study is visualized by ordering the nodes of the causal diagram in two dimensions by their causal order and the time of the observation. Conclusions whether a causal or observational relationship can be estimated from the collected incomplete data can be made directly from the graph.  Causal models with design  offer a systematic and unifying view scientific inference and increase the clarity and speed of communication. Examples on the causal models for a case-control study, a nested case-control study, a clinical trial and a two-stage case-cohort study are presented.

\section{Introduction}
Causal models are commonly used to describe the true or hypothesized causal relationships between a set of variables. The model is typically presented as a directed acyclic graph (DAG), where the nodes represent the variables and the edges represent the causal relationship so that the arrow shows the direction of the effect. A graphical model serves as a tool for visualizing and discussing causal relationships but even more importantly it is a mathematically well-defined object from where causal conclusions can be drawn in a systematic way. Causal calculus \citep{Pearl:1995a,Pearl:book} can be used to estimate causal effects from observational data providing that the study has been carefully designed \citep{Rubin:designtrumps}.

Causal models are not sufficient for the estimation of causal effects without the data. After specifying the causal model and the objectives of the study, the first questions of the researcher should be ``How should the data be collected?'' and ``How should the data collection be taken into account in the analysis?'' \citep{Heckman:sampleselection,Rosenbaum:propensityscore}.
In many fields of science, the data are not obtained as a simple random sample of the population. The pressure of cost-efficiency leads to complex study designs where the expensive measurements are made only for a carefully selected subset of individuals \citep{Reilly:optimalsampling,McNamee:optimaltwostage,Langholz:useofcohort,morgamcasecohort,VanGestel:powerofselectivegenotyping,twostage}. It is therefore crucial to take the study design into account in the estimation of causal effects. The increased complexity of study designs also emphasizes the need for accurate and efficient reporting \citep{STROBE,STROBE_exp,CONSORT,CONSORT_exp}. 

An introduction to causal models with design is given through an example in Section~\ref{sec:basic}. The formal definition of the concept is then presented in Section~\ref{sec:definition}. In Section~\ref{sec:estimation} it is shown how the causal effects can be estimated from a case-control study. Examples describing a clinical trial, a nested case-control study and a two-stage case-cohort study as causal models with design are provided in Section~\ref{sec:examples}. Finally, the benefits, the limitations and the implications of the proposed concept are discussed in Section~\ref{sec:discussion}.

\section{Introductory example} \label{sec:basic}
\begin{figure}
\begin{center}
\includegraphics[width=0.9\columnwidth]{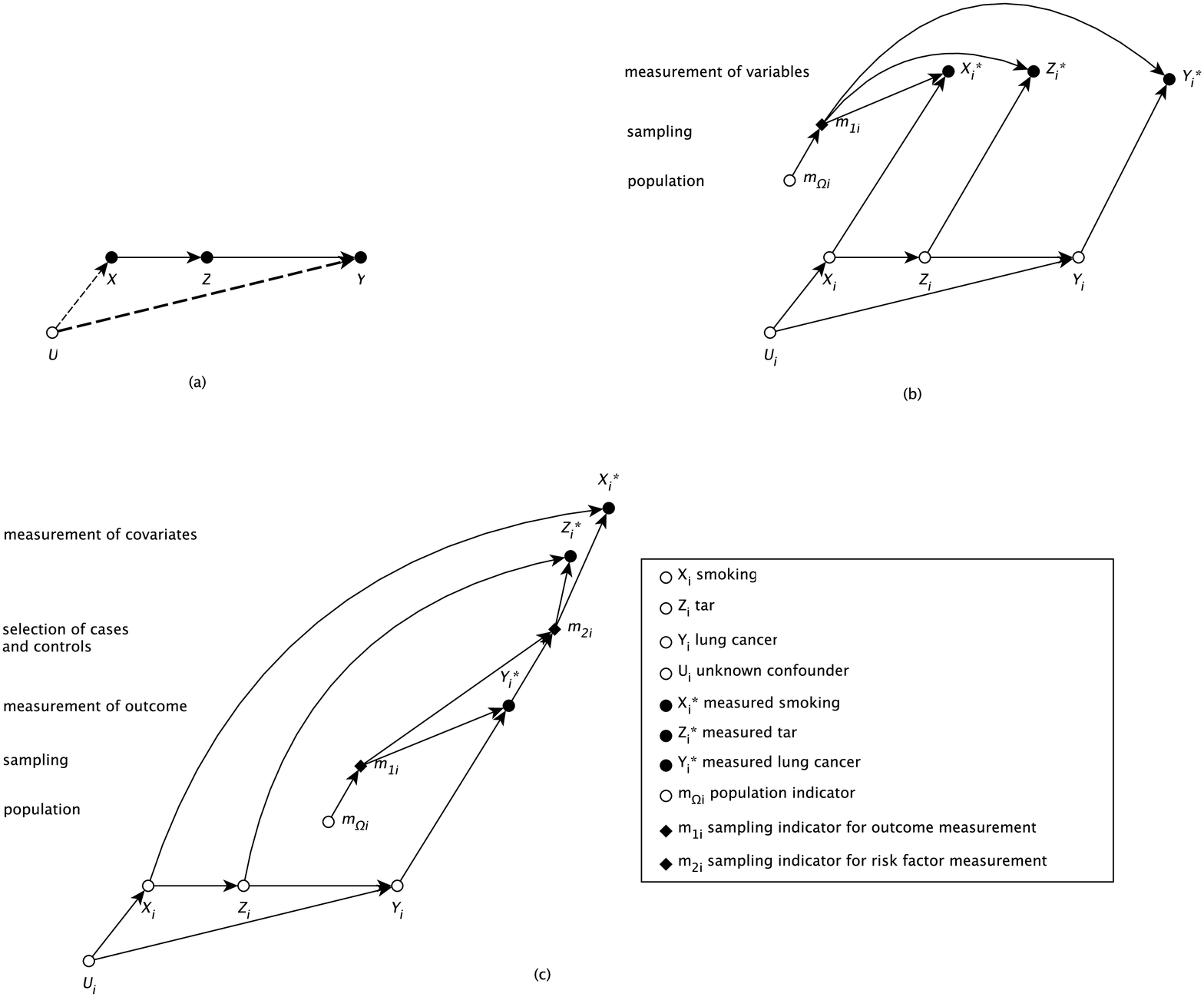}
\caption{Graphical models for the example on the causal effect of smoking to the lung cancer \label{fig:smoking}}
\end{center}
\end{figure}

\citet{Pearl:book} considers an example where the causal effect of smoking $X$ to the lung cancer $Y$ is studied. It is assumed that the causal effect is mediated through the tar deposits in the lungs $Z$. In addition, there might be an unknown confounder $U$ which has a causal effect both to $X$ and $Y$ but not to $Z$. Figure~\ref{fig:smoking}(a) illustrates the causal model. 

In numerical calculations, Pearl implicitly assumes that the data are obtained as a simple random sample from the population. This assumption is made explicit in Figure~\ref{fig:smoking}(b). Variable $m_{\Omega i}$, where subscript $i$ indexes the individuals, represents an indicator for a finite well-defined closed population $\Omega=\{1,\ldots,N\}$. It is defined $m_{\Omega i}=1, i \in \Omega$ and $m_{\Omega i}=0, i \notin \Omega$. Variable $m_{1i}$ represents the sampling. This indicator variable has value 1 if the individual was selected to the sample and 0 otherwise. The arrow from $m_{\Omega i}$ to $m_{1i}$ describes the fact that the sample is selected from the population, i.e. $m_{1i}=1$ implies $m_{\Omega i}=1$. The value of $m_{1i}$ can be determined by the researcher, which is shown in the graph by using diamond symbols for the these nodes.

Variables $X_{i}$, $Z_{i}$ and $Y_{i}$ are related to the underlying population and are not directly observed, which is shown in the visualization with the open circles. Instead, the variables $X^*_{i}$, $Z^*_{i}$ and $Y^*_{i}$ are measured from the sample.  Because these variables are observed, they are shown as filled circles. The value of $Y^{*}_{i}$ is $Y_{i}$ if the individual belongs to the sample, i.e. $m_{1i}=1$; otherwise $Y^{*}_{i}$ is not available. This is described in the graph by arrows from $m_{1i}$ and $Y_{i}$ to $Y^{*}_{i}$. In other words, the causal assumptions, the study design and the data are all presented in the same graph where the causal effects are defined consistently regardless of the type of the variable.

Instead of simple random sampling, case-control designs are often used in epidemiology to study rare diseases. Figure~\ref{fig:smoking}(c) represents a case-control design where the selection for the risk factor measurement is made on the basis of the lung cancer status. In practice, for instance, 1000 lung cancer cases and 1000 non-cases are selected. The lung cancer status $Y^*_{i}$ is determined for the sample $\{i:m_{1i}=1\}$. Smoking $X^*_{i}$ and tar deposits $Z^*_{i}$ are measured for the case-control set $\{i:m_{2i}=1\}$. In the graph, there are arrows from $m_{1i}$ and from $Y^*_{i}$ to $m_{2i}$, which indicates that the selection for case-control set depends on the measured lung cancer status.

It is well known that the study design must be taken into account in the analysis of the data from the case-control design. This means that although Figure~\ref{fig:smoking}(a) presents the causal model for both situations (b) and (c), the analysis of the case-control study (c) differs from the analysis of the simple random sample (b). This difference is made explicit by combining the study design to the causal model. As these causal models with design are causal models, the actual estimation of causal effects can be carried out applying the rules of causal calculus as demonstrated in Section~\ref{sec:estimation}.

\section{Causal models with design} \label{sec:definition}
The formal definition of causal models with design relies on the definition of causal models as presented by \cite{Pearl:book} and the missing data concept presented by \cite{Rubin:inferenceandmissingdata}. The definition of causal models is extended to reflect the elements of inference: the causal assumptions, the study design and the data. The immediate benefit is that the methods of causal calculus are directly applicable for questions related to the study design and estimation. Graphical models with explicit sampling or selection mechanism have been earlier used by \citet{Cooper:bayesian}, \citet{Geneletti:adjusting}, \citet{Didelez:graphical} and \citet{Bareinboim:controllingselectionbias}.

Causal model and probabilistic causal model are defined by \cite{Pearl:book} as follows:
\begin{definition}[Structural Causal Model, Pearl 7.1.1] \label{def:causalmode711l}
 A causal model is a triple $\mathcal{M}=\langle U,V,F \rangle$, where
\begin{enumerate}[(i)]
 \item $U$ is a set of background variables that are determined by factors outside the model;
 \item $V$ is a set $\{V_1,V_2,\ldots,V_n\}$ of variables, called endogenous, that are determined by variables in the model -- that is, variables $U \cup V$; and
 \item $F$ is a set of functions $\{f_1,f_2,\ldots,f_n\}$ such that each $f_i$ is a mapping from (the respective domains of) $U_i \cup PA_i$ to $V_i$ where $U_i \subseteq U$ and $PA_i \subseteq V \setminus V_i$ and the entire set $F$ forms a mapping from $U$ to $V$. In other words, each $f_i$ in $v_i=f(pa_i,u_i)$, $i=1,\ldots,n$, assigns a value to $V_i$ that depends on (the values of) a select set of variables in $V \cup U$, and the entire set $F$ has a unique solution $V(u)$. 
\end{enumerate}
\end{definition}

\begin{definition}[Probabilistic Causal Model, Pearl 7.1.6] \label{def:causalmode716l}
 A probabilistic causal model is a pair $\langle \mathcal{M},P(u) \rangle$ where $\mathcal{M}$ is a causal model and $P(u)$ is a probability function defined over the domain of $U$.
\end{definition}
The causal diagram $G(\mathcal{M})$ of a causal model $\mathcal{M}$ is a directed graph where each node corresponds to a variable and the directed edges point from members of $PA_i$ and $U_i$ toward $V_i$.

%

Causal model with design can be defined as an extension of the probabilistic causal model presented by Pearl where the notation for selection and missing data follows the lines of \citep{Rubin:inferenceandmissingdata}:
\begin{definition}[Causal model with design] \label{def:cmd}
Causal model with design is a probabilistic causal model that fulfills the following conditions:
\begin{enumerate}
\item Each node in the causal diagram is either a causal node, a selection node or a data node. Each node has an information type attribute with possible values: `observed',`not observed', `determined and known' and `determined and unknown'.
\item Each selection node represents a binary variable with the possible values 1 and 0. There is always a unique selection node $M_\Omega$ (population node) which is an ancestor of all selection nodes and has value $M_\Omega=1$.
\item Each data node has two parents, one causal node and one selection node. A causal node cannot be a parent for more than one data node. For a data node $X^*$ with parents causal node $X$ and selection node $M$, it holds
\begin{equation*}
X^*=
\begin{cases}
X, & \mbox{if } M=1 \\
\textrm{NA}, & \mbox{if } M=0
\end{cases}
\end{equation*}
where NA represents a missing value.
\end{enumerate}
\end{definition}

In the first item of Definition~\ref{def:cmd}, the node types are named and the possible values information type attributes are listed. The information type attribute of the variable with the possible values `observed', `not observed' and `determined and known' and `determined and unknown' describes the knowledge of the researcher. In visualizations these types are presented as a filled circle, an open circle, a filled diamond and an open diamond, respectively. In an observational setup, a causal variable $X$ is not observed as such; only the corresponding measurement $X^*$ is observed. In an experimental setup, the values of some causal variables can be determined by the researcher. Usually, causal variables determined by the researcher are known but in principle they can be also unknown if the information on the values set for the variable has been lost after the execution of the experiment. The data are by definition always observed. A selection variable can have all four information types. The value of a selection variable is determined when sampling or other selection is applied to the population. The selection variable can be determined and known or determined and unknown. The latter type, `determined and unknown', may occur, for instance, when the sample is drawn from administrative register with personal identifiers but these are later removed from the data and the researcher is not able to tell which individuals of the population are present in the sample. When the missing data can be identified as an empty record, the selection variable is observed. If the missing individuals are not identified at all, as it is the case in left truncation for instance, the selection variable is not observed.

In the second item of Definition~\ref{def:cmd}, the role of the population and the selection variables is specified. Causal assumptions are always made with respect to some finite population $\Omega$ known as study source in epidemiology \citep{Miettinen:termsandconcepts}. There is always only one population node. If there is more than one conceptual population, the population $\Omega$ can be defined as the union of the conceptual populations. The conceptual population, for instance, a geographical area, becomes a causal variable in the model. If the causal mechanisms differ by the area, the model contains arrows from the area to the causal nodes where the functions $f_j$ differ between the conceptual populations. This allows defining models where some causal relationships are similar across the areas and some are different.  The selection probabilities for the sampling may also differ by the area, which is shown in the model by an arrow from the area to the selection node. 

The members of the population can be a priori known or unknown. In the former case, the researcher has a unique identifier, for instance, the social security number, available for each member of the population before the study. In the latter case, the researcher identifies the members of the population only when they enter to the study. A selection node $M$ induces the subpopulation $\{ i \in \Omega  \mid M_i=1 \}$, which consists of the selected individuals. The causal effects are typically estimated for the population $\Omega$ but, for instance, in epidemiological cohort studies the effects are often estimated only for the cohort $\{ i \in \Omega  \mid M_i=1 \}$, also known as study base \citep{Miettinen:termsandconcepts}.

In the third item of Definition~\ref{def:cmd}, the relations of the causal variables, the selection variables and the data are specified. The value of random variable $X_i$ is measured only if the individual $i$ is selected to be measured, which is indicated by the selection variable $M_i$. This means that the measured value $X_i^*$ is a random variable which depends on the variables $X_i$ and $M_i$ in a deterministic way. The definition of a univariate random variable is extended so that in addition to real axis, a random variable may also have a special value `NA' which indicates missing data. With this definition, all elements of scientific inference can be expressed as random variables and their causal relationships.  If a data node or a selection node has a directed path to a causal node, the measurement or the selection has a causal effect to the underlying causal variable. This may be the case, for instance, in health examination studies where the participation to the study may increase the awareness on the healthy life 
style and consequently also have an impact to the later measurements of health indicators.

In a causal model, the causal effects define a partial ordering between the variables. In addition to this causal time, the time of observation can be linked to each variable in a causal model with design. Together the causal time and the observational time define the relative location of each node in a visualization where the causal time is presented on x-axis and the observational time on y-axis. To make the visualization more informative, the stages of the study can be used as labels for the y-axis as it is done in the examples of Sections~\ref{sec:basic} and \ref{sec:examples}.

Measurement error can be added to a causal model with design by introducing two causal variables: the original variable $X_i$ and the version with measurement error $\tilde{X}_i$. In the graph there is an arrow from $X_i$ to $\tilde{X}_i$. Both $X_i$ and $\tilde{X}_i$ are unobserved and only $\tilde{X}_i^*$ is observed for the sample. Variable $X_i^*$ is usually unobserved unless some kind of benchmark measurements without measurement error are carried out for a subsample. If two variables $X_i$ and $Y_i$ have correlated measurement errors, an explicit unobserved causal variable $U$ is needed to describe the structure of the measurement error. In the graph, there are arrows from $U$ to $\tilde{X}_i$ and to $\tilde{Y}_i$ in addition to arrows $X_i \rightarrow \tilde{X}_i$ and $Y_i \rightarrow \tilde{Y}_i$.  Again only $\tilde{X}_i^*$ and $\tilde{Y}_i^*$ are observed for the sample.



In causal models with design, sampling and nonresponse are formally treated in a similar way; the only difference is the type of the selection node which is `determined' for sampling and `observed' for nonresponse. Some conclusions on the type of missing data mechanism \cite{Rubin:inferenceandmissingdata} can be made directly from the causal model with design.  Let $M$ to be the selection variable for the measurement $Y^{*}$ of causal variable $Y$. If there is no (undirected) path from $Y$ to $M$ except through $Y^*$, the data on $Y$ are missing completely at random (MCAR), more precisely everywhere MCAR \citep{Seaman:MAR}. If there is an arrow from $Y$ to $M$, the data are missing not at random (MNAR). The traditional MCAR/MAR/MNAR classification concerns the data as a whole whereas causal models with design provide a description of the missingness mechanism variable by variable. 

Many recent theoretical result on missing data and selection bias in causal inference can be applied to causal models with design. As these results are not defined directly for causal models with design but for other extensions of causal models, transformations are applied as the first step. \citet{Mohan:missingdata} consider estimation when data are MNAR and  derive conditions a ``missingness graph'' should satisfy to ensure the existence of a consistent estimator for a given probabilistic relation. In order to utilize these results, a causal model with design can be collapsed to a missingness graph by removing the intermediate selection nodes, i.e. selection nodes that are not parents of a data node. Formally this can be defined as follows:
\begin{definition}[Collapse to a Missingness Graph]
Missingness graph $H$ is a collapse of causal model with design $\mathcal{M}$ with causal diagram $G(\mathcal{M})$ if (i) the set of nodes in $H$ consists of the causal nodes of $\mathcal{M}$, the data nodes of $\mathcal{M}$ and such selection nodes of $\mathcal{M}$ that are parents of some data node, (ii) there exist an edge from node $X$ to node $Y$ in $H$ if there exists an edge from $X$ to $Y$ in $G(\mathcal{M})$ or if $X$ is a causal node and $Y$ is a selection node and there exists a directed path from $X$ to $Y$ in $G(\mathcal{M})$.
\end{definition}

The results and algorithms by \citet{Bareinboim:controllingselectionbias} can be used to mitigate and sometimes to eliminate the selection bias caused by preferential data collection. The results are applicable in the important special case where a single selection node (often marked by $S$) is the parent for all data nodes. In order to apply these results, a causal model with design is first collapsed to a missingness graph and then the data nodes are removed. The transformed graph contains the selection node $S$ and all causal nodes. The results by \citet{Didelez:graphical}, \citet{Geneletti:adjusting} and \citet{Cooper:bayesian} can be also applied to the same transformed graph.

\citet{bareinboim2013general,Bareinboim:metatransportability} consider theoretical conditions for the transfer of experimental results from one or several populations to other populations. Causal models with design have only one population but the transportability results can be used between the conceptual populations. The application of the results and the algorithms by \citet{bareinboim2013general,Bareinboim:metatransportability} requires that the causal model with design has been collapsed to a selection diagram as follows:
\begin{definition}[Collapse to a Selection Diagram for Transportability]
Selection diagram $H_S$ is a collapse of causal model with design $\mathcal{M}$ with respect to a set of selection variables $S$ if (i) the conceptual populations of $\mathcal{M}$ are identified by the variables of $S$ (ii) the set of nodes in $H_S$ consist of the causal nodes of $\mathcal{M}$ (iii) there exist an edge from node $X$ to node $Y$ in $H_S$ if there exists an edge from $X$ to $Y$ in $G(\mathcal{M})$ and $Y$ does not belong to $S$.
\end{definition}
Other recent developments that can be applied to causal models with design include the results on the testability of counterfactuals \citep{Shpitser:counterfactuals} and z-identifiability of surrogate experiments \citep{Bareinboim:zidentifiability}.



\section{Estimation of causal effects} \label{sec:estimation}
The following steps are required to estimate causal effects using causal models with design:
\begin{enumerate}
 \item Specify the causal model.
 \item Check the identifiability of the causal effect in the causal model using the results by \citet{tian2002general}, \citet{shpitser2006identificationjoint,shpitser2006identificationconditional} and \citet{Bareinboim:zidentifiability}.  If the effect can be identified, use the rules of causal calculus \citep{Pearl:1995a,Pearl:book} to express the causal effect in terms of observed probabilities.
 \item Expand the causal model to the causal model with design.
 \item Form the likelihood according to the causal model with design and integrate it over the unobserved variables.
 \item Estimate the parameters needed to calculate the causal effect as derived in Step 2.
\end{enumerate}

Causal models with design allow the estimation of causal effects in complex designs using only the rules of causal calculus and the likelihood. This requires, however, that the causal effect can be expressed in terms of observed probabilities (Step 2) and the parameters of the likelihood can be estimated (Step 5).  Even if a causal effect is not identifiable in the general nonparametric form it may still be identifiable under a specific parametric model. For example, an instrumental variable may help to identify a causal effect in a linear model but not in a nonlinear model \citep{Pearl:book} and the average causal effect in clinical trials with noncompliance can be identified under specific assumptions \citep{angrist1996identification}. Even if a causal effect is identifiable in the general nonparametric form, it may not be estimable from the collected data. A well-known example is the MNAR situation where a variable has a causal effect on its selection variable and the estimation is not possible in general without strong assumptions on the selection mechanism \citep{littlerubin2002}.   



As an example of the estimation procedure, the smoking and lung cancer example of Section~\ref{sec:basic} is considered again. The causal model is specified in Figure~\ref{fig:smoking}(a) (Step 1). The goal is the estimate the causal effect $p(y \mid \doo(X=x))$ where the do-operator represents action/intervention. The result (Step 2)
\begin{equation} \label{eq:frontdoor}
 p(y \mid \doo(X=x)) = \sum_{z} p(z \mid X=x) \sum_{x'}p(y \mid X=x',Z=z) p(X=x')
\end{equation}
is obtained applying the following three rules of causal calculus \citep{Pearl:1995a,Pearl:book}:
\begin{enumerate}
 \item Insertion and deletion of observations:
 \begin{align*}
  & p(y \mid \doo(x),z,w) = p(y \mid \doo(x),w), \\ \nonumber
   & \textrm{ if }  (Y \independent Z \mid X,W) \textrm{ in the graph } G_{\overline{X}}.
\end{align*} 
 \item Exchange of action and observation:
 \begin{align*}
  & p(y \mid \doo(x),\doo(z),w) = p(y \mid \doo(x), z, w), \\ \nonumber
  & \textrm{ if }  (Y \independent Z \mid X,W) \textrm{ in the graph } G_{\overline{X}\underline{Z}}.
 \end{align*}
\item Insertion and deletion of actions:
\begin{align*}
 & p(y \mid \doo(x),\doo(z),w) = p(y \mid \doo(x), w), \\ \nonumber
 & \textrm{ if }  (Y \independent Z \mid X,W) \textrm{ in the graph }  G_{\overline{X}\overline{Z(W)}},
\end{align*}
where $Z(W)$ is the set of the $Z$-nodes that are not ancestors of any $W$-node in the graph $G_{\overline{X}}$.
\end{enumerate}
Here $G_{\overline{X}}$ represents a graph where the incoming edges of the set of nodes $X$ are removed, $G_{\underline{X}}$ represents a graph where the outgoing edges of the set of nodes $X$ are removed and $G_{\overline{X}\underline{Z}}$ represents a graph where the incoming edges of the $X$-nodes and the outgoing edges of the $Z$-nodes are removed. The rules of causal calculus are sufficient for deriving all identifiable causal effects from observational data \citep{Pearlsiscomplete,shpitser2006identificationjoint} and experimental data \citep{Bareinboim:zidentifiability} for a given population.  Alternatively, the back-door and front-door criteria \citep{Pearl:book} and the moralization \citep{Lauritzen:independence} can be also used to derive formulas for the causal effects. Algorithms for the automated application of causal calculus have been developed \citep{tian2002general,Pearlsiscomplete,shpitser2006identificationjoint,Bareinboim:zidentifiability}.
 
Next consider the case-control design of Figure~\ref{fig:smoking}(c) (Step 3). To estimate the causal effects, the model parameters must be estimated from the data collected according to this design. The likelihood can be factorized according to the graphical model 
\begin{align*} 
& p( \mathbf{m}_{\Omega},\mathbf{m}_1,\mathbf{m}_2,\mathbf{Z},\mathbf{X},\mathbf{Y},\mathbf{U},\mathbf{Z}^*,\mathbf{X}^*,\mathbf{Y}^* \mid \boldsymbol{\theta},\boldsymbol{\psi})= \\ 
& \prod_{i=1}^{N} p(m_{\Omega i}) p(m_{1i} \mid m_{\Omega i},\boldsymbol{\psi}) p(U_i) p(X_i \mid U_i,\boldsymbol{\theta}) p(Z_i \mid X_i, \boldsymbol{\theta})
 p(Y_i \mid Z_i,U_i,\boldsymbol{\theta})\\
& \times  p(m_{2i} \mid m_{1i},Y_i,\boldsymbol{\psi})= 
\\
 & \prod_{\{i:m_{2i}=1 \}} p(m_{1i}=1 \mid m_{\Omega i},\boldsymbol{\psi}) p(U_i) p_X(X_i^* \mid U_i,\boldsymbol{\theta}) 
 p_Z(Z_i^* \mid X_i=X_i^*,\boldsymbol{\theta})  p_Y(Y_i^* \mid Z_i=Z_i^*,U_i,\boldsymbol{\theta}) \\
 & \times p(m_{2i}=1 \mid m_{1i}=1,Y_i^*,\boldsymbol{\psi})
 \\
  & \prod_{\{i:m_{2i}=0,m_{1i}=1 \}} p(m_{1i}=1 \mid m_{\Omega i},\boldsymbol{\psi}) p(U_i) p(X_i \mid U_i,\boldsymbol{\theta}) 
 p(Z_i \mid X_i,\boldsymbol{\theta})  p_Y(Y_i^* \mid Z_i,U_i,\boldsymbol{\theta}) \\
 & \times p(m_{2i}=0 \mid m_{1i}=1,Y_i^*,\boldsymbol{\psi})
 \\
   & \prod_{\{i:m_{1i}=0 \}} p(m_{1i}=0 \mid m_{\Omega i},\boldsymbol{\psi}) p(U_i) p(X_i \mid U_i,\boldsymbol{\theta}) 
 p(Z_i \mid X_i,\boldsymbol{\theta})  p_Y(Y_i \mid Z_i,U_i,\boldsymbol{\theta}),
\end{align*} 
where $\boldsymbol{\theta}$ represents the model parameters, $\boldsymbol{\psi}$ represents parameters related to the design and the vector notation, such as $\mathbf{m}_1=(m_{11},\ldots,m_{1N})^T$, refers to the variables for all individuals $\{1,\ldots,N\}$ in the population. The distributions are defined with respect to the first argument unless otherwise specified. The likelihood of the observed data is obtained as an integral over the unknown variables $\mathbf{Z}$, $\mathbf{X}$, $\mathbf{Y}$ and $\mathbf{U}$ (Step 4) 
\begin{align*} 
& p( \mathbf{m}_{\Omega},\mathbf{m}_1,\mathbf{m}_2,\mathbf{Z}^*,\mathbf{Y}^*,\mathbf{X}^* \mid \boldsymbol{\theta},\boldsymbol{\psi})=  \\ 
 & \prod_{\{i:m_{2i}=1 \}} p(m_{1i}=1 \mid m_{\Omega i},\boldsymbol{\psi}) p_X(X_i^* \mid \boldsymbol{\theta}) 
 p_Z(Z_i^* \mid X_i=X_i^*,\boldsymbol{\theta}) p_Y(Y_i^* \mid Z_i=Z_i^*,X_i=X_i^*,\boldsymbol{\theta}) \\ 
 & \times p(m_{2i}=1 \mid m_{1i}=1,Y_i^*,\boldsymbol{\psi})
 \\ 
  & \prod_{\{i:m_{2i}=0,m_{1i}=1 \}} p(m_{1i}=1 \mid m_{\Omega i},\boldsymbol{\psi}) p_Y(Y_i^* \mid \boldsymbol{\theta}) \\ 
 & \times p(m_{2i}=0 \mid m_{1i}=1,Y_i^*,\boldsymbol{\psi})
 \\ 
  & \prod_{\{i:m_{1i}=0 \}} p(m_{1i}=0 \mid m_{\Omega i},\boldsymbol{\psi}). \numberthis \label{eq:smokinglikelihood2}
\end{align*} 
As the selection $m_1$ is random sampling from the population, the term $p(m_{1i}=0 \mid m_{\Omega i},\boldsymbol{\psi})$ may be ignored in the estimation of $\boldsymbol{\theta}$. The selection $m_2$ depends on the response $Y$ and the term $p(m_{2i}=0 \mid m_{1i}=1,Y_i^*,\boldsymbol{\psi})$ must not be ignored. Note also that although $X$ is not a parent of $Y$ in the causal model, the likelihood~\eref{eq:smokinglikelihood2} has the term $p(Y=1 \mid X=x,Z=z)$.

In Step 5 the likelihood must be written in a parametric form. Finding a good parametrization, i.e. finding a good statistical model, is purely a statistical problem. Causal considerations are not needed in the model selection or in the parameter estimation and the vast literature on these topics is directly applicable. It follows from equation~\eref{eq:frontdoor} that the probabilities $p(x)$, $p(z \mid X=x)$ and $p(y \mid X=x,Z=z)$ are needed to estimate $p(y \mid \doo(X=x))$. The same probabilities are also components in the likelihood~\eref{eq:smokinglikelihood2} and it is therefore natural to parametrize them. For simplicity \citet{Pearl:book} assumes that the variables $X$, $Z$ and $Y$ have possible values 0 and 1.  The observed probabilities mentioned above can be now parametrized as follows:
\begin{align*} 
 & p(X=1)=\theta_X, \\
 & p(Z=1 \mid X=x) = \theta_Z + x \theta_{ZX}, \\
 & p(Y=1 \mid X=x,Z=z) = \theta_Y + x \theta_{YX} + z \theta_{YZ} + xz\theta_{YZX}. 
\end{align*}
With this parametrization, the causal effect of smoking to the risk of lung cancer given by equation~\eref{eq:frontdoor} can be written as
\begin{align} 
  p(Y=1 \mid \textrm{do}(X=1)) = 
 & (\theta_Z+\theta_{ZX}) \big( \theta_X(\theta_Y+\theta_{YX}+\theta_{YZ}+\theta_{YZX}) + \nonumber \\ & (1-\theta_X) (\theta_Y+\theta_{YZ}) \big) + \nonumber \\
  & (1-\theta_Z-\theta_{ZX}) \big( \theta_X(\theta_Y+\theta_{YX})+(1-\theta_X)\theta_Y\big) \label{eq:paramcausaleffect1}  \\ \nonumber
   p(Y=1 \mid \textrm{do}(X=0)) = & \theta_Z \big( \theta_X(\theta_Y+\theta_{YX}+\theta_{YZ}+\theta_{YZX}) + (1-\theta_X)(\theta_Y+\theta_{YZ}) \big) +  \\
  & (1-\theta_Z) \big( \theta_X(\theta_Y+\theta_{YX})+(1-\theta_X)\theta_Y\big). \label{eq:paramcausaleffect2}
 \end{align}
These equations link the model parameters $\boldsymbol{\theta}=(\theta_X,\theta_Z,\theta_Y,\theta_{ZX},\theta_{YX}, \theta_{YZ},\theta_{YZX})$ to the causal effects. 
The dependency of the selection probability on $Y^*$ may be parametrized as 
\begin{equation*}
 p(m_2=1 \mid Y^*=y) = \psi + y\psi_Y.
\end{equation*}


As the variables are binary, the data collected according to the case-control design can be presented in the form of frequencies given in Table~\ref{tab:ndata}. The size of the population is $N=N_{11}+N_{10}+N_{01}+N_{01}$ where $N_{11}$ is the number of cases selected, $N_{10}$ is the number of non-cases selected, $N_{01}$ is the number of cases not selected and $N_{00}$ is the number of non-cases not selected. In the other words, it is assumed that the lung cancer prevalence in the population is known. The log-likelihood derived from the likelihood~\eref{eq:smokinglikelihood2} becomes
\begin{align*}
& n_{1\cdot\cdot}\log \theta_X + n_{0\cdot\cdot}\log(1-\theta_X) + n_{11\cdot}\log(\theta_Z+\theta_{ZX}) + + n_{01\cdot}\log(\theta_Z) +\\
& n_{10\cdot}\log(1-\theta_Z-\theta_{ZX}) + n_{00\cdot}\log(1-\theta_Z) + \\
& n_{111}\log(\theta_Y+\theta_{YX}+\theta_{YZ}+\theta_{YZX}) + n_{101}\log(\theta_Y+\theta_{YX}) + n_{101}\log(\theta_Y+\theta_{YZ}) + \\
& n_{001}\log(\theta_Y) + n_{110}\log(1-\theta_Y-\theta_{YX}-\theta_{YZ}-\theta_{YZX}) +\\
& n_{100}\log(1-\theta_Y-\theta_{YX}) + n_{010}\log(1-\theta_Y-\theta_{YZ}) + n_{000}\log(1-\theta_Y) + \\
& N_{01}\log(\theta_Y') + N_{00}\log(1-\theta_Y')+\\
& N_{11}\log(\psi+\psi_Y)+N_{10}\log(\psi)+N_{01}\log(1-\psi-\psi_Y)+N_{00}\log(1-\psi),
\end{align*}
where $\cdot$ represents summation over the corresponding marginal and 
\begin{align*}
 \theta_Y' = p(Y=1) = & (1-\theta_X)(1-\theta_Z)\theta_Y+\theta_X(1-\theta_Z-\theta_{ZX})(\theta_Y+\theta_{YX}) + \\
 & (1-\theta_X)\theta_Z(\theta_Y+\theta_{YZ})+\theta_X(\theta_Z+\theta_{ZX})(\theta_Y+\theta_{YX}+\theta_{YZ}+\theta_{YZX})
\end{align*}
is a shorthand notation for the marginal probability of $Y$. The maximum likelihood estimates of $\boldsymbol{\theta}$ can be obtained by numerical optimization of the log-likelihood. Naturally, a Bayesian analysis may be carried out as well.

\begin{table}[htb]
\caption{Data collected from the case-control study} 
\label{tab:ndata}
\begin{center}
\begin{tabular}{lcc|cc}
& \multicolumn{2}{c}{Notation} \vline& \multicolumn{2}{c}{Numerical illustration}\\
 & $Y=1$ & $Y=0$ & $Y=1$ & $Y=0$\\
\hline
 $X=0,Z=0$ & $n_{001}$ & $n_{000}$ & 100 & 814  \\
  $X=1,Z=0$ & $n_{101}$ & $n_{100}$ & 47 & 5\\
   $X=0,Z=1$ & $n_{011}$ & $n_{010}$ & 3 & 45\\
    $X=1,Z=1$ & $n_{111}$ & $n_{110}$ & 850 & 136\\
 \hline 
 sum & $N_{11}$ & $N_{10}$ & 1000 & 1000 
\end{tabular}
\end{center}
\end{table}

For a numerical illustration, consider a case-control study where 1000 lung cancer cases and 1000 controls are selected for the covariate measurements. The parameters $\boldsymbol{\theta}$ are set according to the (unrealistic) population probabilities used in \citep[page 84]{Pearl:book}. The expected frequencies are shown in Table~\ref{tab:ndata}. With these frequencies and the numbers of non-selected individuals $N_{01}=8500$ and $N_{00}=9500$, the maximum likelihood estimates $\hat{\theta}_X=0.50$, $\hat{\theta}_Y=0.10$, $\hat{\theta}_Z=0.050$, $\hat{\theta}_{ZX}=0.90$, $\hat{\theta}_{YZ}=-0.043$, $\hat{\theta}_{YX}=0.79$, $\hat{\theta}_{YZX}=-0.0019$, $\hat{\psi}=0.095$, $\hat{\psi}_Y=0.010$ and $\hat{\theta}_Y'=0.48$ are obtained. The equations~\eref{eq:paramcausaleffect1} and \eref{eq:paramcausaleffect2} give the causal effects
\begin{align*}
 p(Y=1 \mid \textrm{do}(X=1)) = 0.456 \\
  p(Y=1 \mid \textrm{do}(X=0)) = 0.495,
\end{align*}
which are similar to the causal effects estimated from the whole population in \citep[page 84]{Pearl:book}. The differences in the third decimal are due to the rounding of the expected frequencies in Table~\ref{tab:ndata} to the nearest integer.

\section{Examples with complex study design} \label{sec:examples}
The examples presented in this section aim to demonstrate how causal models with design can describe the essential features of complex experimental and observational studies in a precise and illustrative way. The examples are from medicine and epidemiology where complex study designs are commonly used. The first example is based on a real study and causal models with design are used to make conclusions on the identifiability of various causal effects from data missing not at random. The two other examples describe imaginary but realistic scenarios.  

Causal graphs with design remove the ambiguity related to the common names of study designs such retrospective study, prospective study, cohort study, case-control study and two-stage study \citep{Vandenbroucke:retroprospect,Knol:whatdocasecontrol}. The process of the data collection can be seen directly from the causal graph with design. 

For the estimation of causal effects, the procedure presented in Section~\ref{sec:estimation} is applicable. Causal models with design are also useful in the estimation of predictive models when the study design and the missing data mechanism must be taken into account in the analysis. The likelihood factorized according to the causal model with design offers a natural starting point for the parameter estimation in both the frequentist and the Bayesian approach. The idea is to write first the full likelihood for the data, the design and the latent variables, and then see which parts of the likelihood are not needed in the estimation of the parameters of the interest. The likelihood functions for the examples of this section are given in Appendix. 

Figure~\ref{fig:morgamcasecohort} illustrates a causal model with design for the two-stage case-cohort design used in the MORGAM Project \citep{morgamcasecohort,evans:2005}. The project aims to estimate the impact of classic and genetic risk factors to the risk of cardiovascular diseases.  Currently 15 cohorts from 6 countries participate in the genetic component of the project. Most of the cohorts are selected as random samples of the underlying population of certain age range, typically 24--65 years although there is variation between the cohorts. Over 50,000 individuals have been examined for the classic risk factors and followed up for mortality and disease endpoints. Due to the cost of genotyping, genes have been measured only for a subset of each cohort. Over 10,000 individuals have been genotyped in the case-cohort setting.

The causal assumptions are described using four variables: genetic risk factors $Z_i$, classic risk factors $X_i$ and health status at baseline $Y_{0i}$ and at the end of the follow-up $Y_i$. Here classic risk factors are understood to include the actual risk factors such as smoking, cholesterol and blood pressure as well as all relevant background variables measured at baseline. The internal causal structure between these variables is not specified because it is not needed in the following considerations. From the graph it can be read that genes may affect the disease risk directly and via classic risk factors. Classic risk factors measured at baseline may be affected by the health status at baseline. The following conclusions can be made using causal calculus:  
\begin{enumerate}
\item The causal effect of genes to disease corresponds to the observed effect in the population 
\begin{equation*}
 p(Y \mid \textrm{do}(Z=z)) = p(Y \mid Z=z).
\end{equation*}
\item The causal effect of classic risk factors to disease in the population is confounded by genes and health status at baseline
\begin{equation} \label{eq:morgambackdoor}
 p(Y \mid \textrm{do}(X=x)) = \int\! \int p(Y \mid X=x,Z=z,Y_0=y_0)p(Z=z,Y_0=y_0) \dd z \dd y_0.
\end{equation}
\item The causal effect of classic risk factors to disease conditioned on health status at baseline is confounded by genes
\begin{equation} \label{eq:morgamhealthybackdoor}
 p(Y \mid \textrm{do}(X=x), Y_0=y_0) =  \int p(Y \mid X=x,Z=z,Y_0=y_0)p(Z=z \mid Y_0=y_0) \dd z .
\end{equation}
\item Causal effect of genes, classic risk factor and health status at baseline corresponds to the observed conditional effect in the population 
\begin{align*}
  & p(Y \mid \textrm{do}(Z=z),\textrm{do}(Y_0=y_0),\textrm{do}(X=x)) = \\  
  & p(Y \mid Z=z,\textrm{do}(Y_0=y_0),\textrm{do}(X=x)) = \\
  &  p(Y \mid Z=z,Y_0=y_0,\textrm{do}(X=x))=p(Y \mid Z=z,Y_0=y_0,X=x).
\end{align*}
\end{enumerate}

In order to see whether these effects can be estimated from the collected data, the study design need to be investigated. The population is defined to include all individuals living in a specified geographical area and born in specified years. However, the sampling frame is not the birth cohort but the individuals alive at baseline. In other words, individuals who have died before baseline are left truncated. In the graph this left truncation is shown by an arrow from $Y_{0i}$ to the unobserved selection node $M_{0i}$. Sampling, denoted by node $m_{1i}$, is carried out and each selected individual makes a decision on the participation $M_{1i}$. This decision depends on health status $Y_{0i}$ , socio-economic status  and classic risk factors $X_i$ \citep{chou1997characteristics,hara2002comparison,cohen2002nonrespondents,jousilahti2005total,drivsholm2006representativeness,knudsen2010health,alkerwi2010comparison}.  The data are MNAR because the fact whether $X_i$ and $Y_{0i}$ are measured depend on the values of these variables. However, the missingness mechanism may still be ignorable in some analyses. Applying d-separation to causal model with design it can be concluded that
\begin{align}
& M_{1i} \not \independent Y_{i} \mid Z_{i} \label{eq:morgamnoz} \\
& M_{1i} \not \independent Y_{i} \mid Z_{i}, Y_{0i} \label{eq:morgamnozcondy0} \\
&  M_{1i} \independent Y_{i} \mid X_{i}, Y_{0i} \label{eq:morgamcondx}\\
&   M_{1i} \independent Y_{i} \mid Z_{i}, X_{i}, Y_{0i} \label{eq:morgamcondz}.
\end{align}
From result~\eref{eq:morgamnoz} it follows that the cohort data (and consequently the case-cohort data) cannot be used to estimate the causal (or predictive) effect genetic risk factors to disease in the population without accounting for the missingness mechanism. Result~\eref{eq:morgamnozcondy0} tells that conditioning on the health status at baseline does not change the situation qualitatively. From result~\eref{eq:morgamcondx} it follows that the cohort data can be used to estimate the predictive model $p(Y_i \mid X_i=x_i,Y_{0i}=y_{0i})$ for the healthy population. This kind of conditioning on the health status is commonly used in epidemiology and has been applied also in the MORGAM Project, e.g. in \citep{stroke}. To estimate the causal effects of classic risk factors in the population, the missingness mechanism must be taken into account because equation~\eref{eq:morgambackdoor} contains the distribution $p(Z=z,Y_0=y_0)$, which is potentially different for participants and non-participants. Similarly, the term $p(Z=z \mid Y_0=y_0)$ in equation~\eref{eq:morgamhealthybackdoor} implies that the missingness mechanism must be taken into account also when the causal effects of classic risk factors are estimated for the healthy population. From result~\eref{eq:morgamcondz} it follows that the case-cohort data can be used to estimate the causal effect of genetic risk factors for the healthy population on the condition of classic risk factors. As the case-cohort selection $M_{2i}$ depends on $Y$, the case-cohort selection should be taken into account in the estimation \citep{morgamcasecohort,kulathinal:2006}.

The data on health status at baseline $Y_{0i}^*$ includes information on non-fatal cardiovascular events before baseline. Restricting the analysis to the individuals healthy at baseline, i.e. removing individuals with prior non-fatal events, discards a considerable amount of potentially useful data. In the MORGAM Project, several attempts have been made to use these so called baseline cases. In \citep{newlyidentified} baseline cases are analyzed separately. The joint analysis of baseline cases and follow-up cases requires compensation for the left truncation, which can be done using nonparametric imputation \citep{npmi} or conditional likelihood \citep{prevalenceincidence}. These works, however, do not take the non-participation into account.

\begin{figure}
 \begin{center}
\includegraphics[width=0.9\columnwidth]{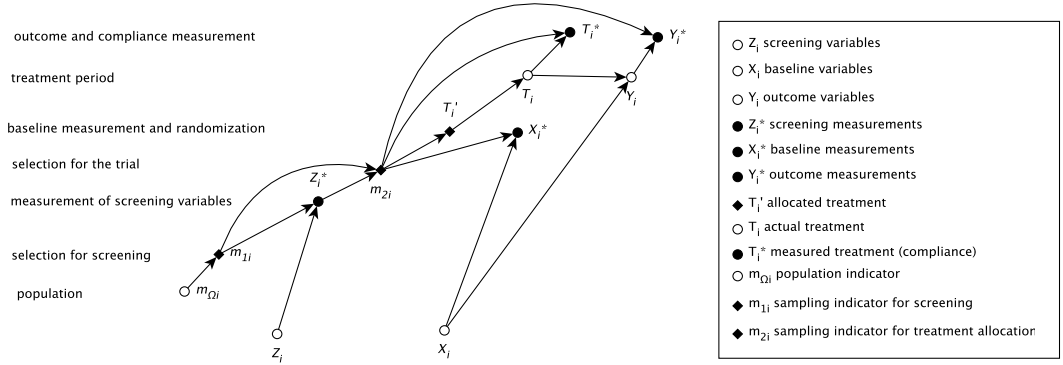}
\caption{Causal model with design for a clinical trial. A sample $\{i:m_{1i}=1\}$ is selected for screening from the population $\{i:m_{\Omega i}=1\}$. The inclusion for the trial $m_{2i}$ is based on the screening variable $Z_{i}^{*}$. At the baseline, covariate $X_i^{*}$ is measured for the trial participants and a randomized decision on the treatment $T_{i}'$ is made. The actual treatment $T_{i}$ during the treatment period may differ from the intended treatment $T_{i}'$  because of non-compliance. The outcome $Y_{i}$ depends on the covariate $X_{i}$ and the treatment $T_{i}$. At the end of the treatment period, measurements for the observed outcome $Y_{i}^{*}$ and the observed treatment $T_{i}^{*}$ are made.  \label{fig:clinicaltrial}}
\end{center}
\end{figure}

Figure~\ref{fig:clinicaltrial} shows how the experimental setup of a clinical trial can be described in a causal model with design. The treatment in the clinical trial is a causal variable determined by the researcher by the means of randomization. In the graph, this is presented by causal node $T_{i}'$ which has the type determined and known. The example also demonstrates the compliance problem encountered in clinical trials: the actual treatment may differ from the allocated treatment if the participant does not follow the instructions given. In the graph, there is an arrow from $T_{i}'$ to the actual treatment $T_i$ and $T_{i}'$ affects outcome $Y_i$ only through $T_i$. In the intention-to-treat analysis, the observed outcome $Y_{i}^{*}$ is explained by the intended treatment $T_{i}'$ using all included participants in the trial $\{i:m_{2i}=1\}$. In the per-protocol analysis, only the compliant participants with $T_{i}'=T_{i}^{*}$ are included.

\begin{figure}
\begin{center}
\includegraphics[width=0.9\columnwidth]{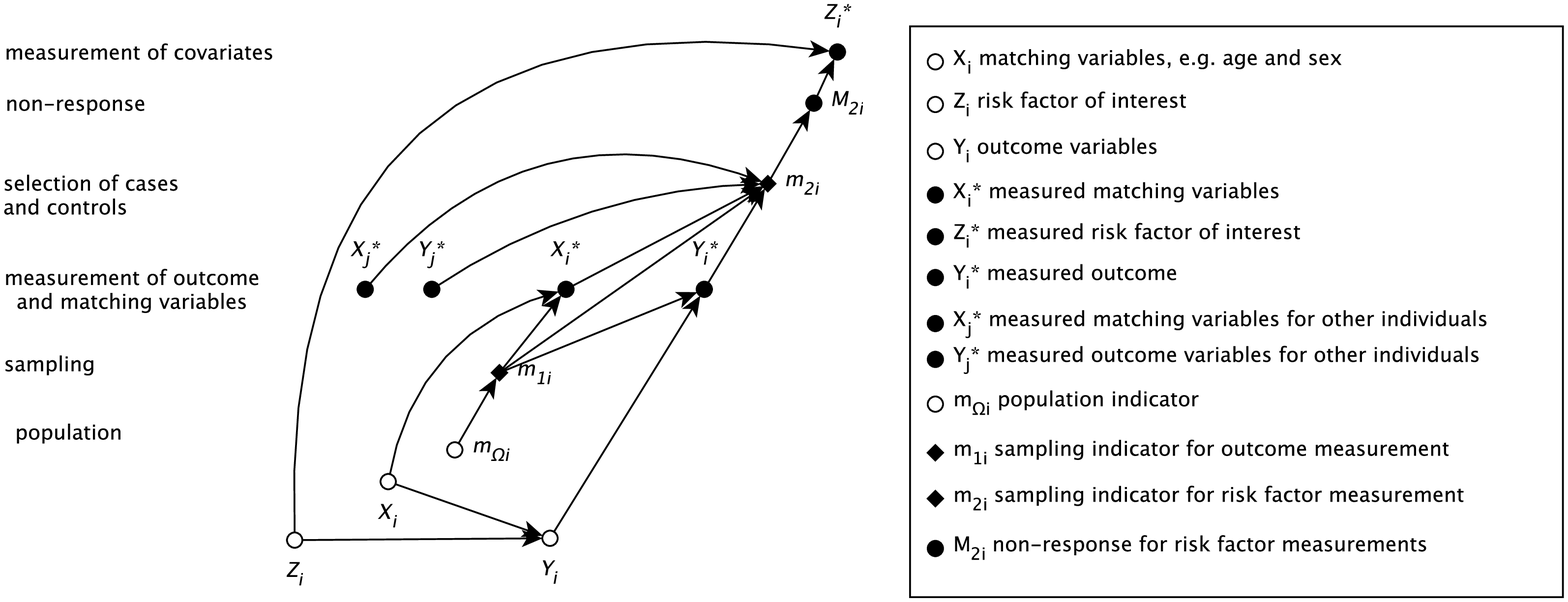}
\caption{Causal model with design for a nested case-control study. The idea of the case-control design is to select the individuals for the measurement of the expensive risk factor $Z_i$ on the basis the outcome $Y_i$ and the inexpensive risk factor $X_i$. At the first stage, a sample $\{i:m_{1i}=1\}$ is selected from the population $\{i:m_{\Omega i}=1\}$ and variables $X_i^*$ and $Y_i^*$ are measured. The selection of cases and controls 
$m_{2i}$ depends not only on measurements of individual $i$, $X_i^*$ and $Y_i^*$, but also on the outcome $Y_j^*$ and the covariate $X_j^*$ of all other individuals in the sample. Each individual has a similar causal graph which has been omitted in the figure. The nonresponse $M_{2i}$ reflects the fact the measurement $Z_{i}^{*}$ may not be available for all individuals selected to the case-control set. \label{fig:nestedcc}}
\end{center}
\end{figure}

Figure~\ref{fig:nestedcc} illustrates a situation where there is a dependence structure between the selection variables of the individuals in the sample. In a nested case-control design, the controls are selected considering the individuals at risk at the time (age or calendar time) of the disease event. A control may later become a case which creates a complicated dependence structure between the selection probabilities  \citep{secondary}. Consequently, the selection probability for individual $i$ depends on the covariates and outcomes of all other individuals. In the graphical presentation drawn for individual $i$,  the case-control selection node $m_{2i}$ has incoming arrows from $X_i^*$, $Y_i$, $X_j$ and $Y_j$ where index $j$ is used to refer to all other individuals. 

\begin{figure}
\begin{center}
\includegraphics[width=0.9\columnwidth]{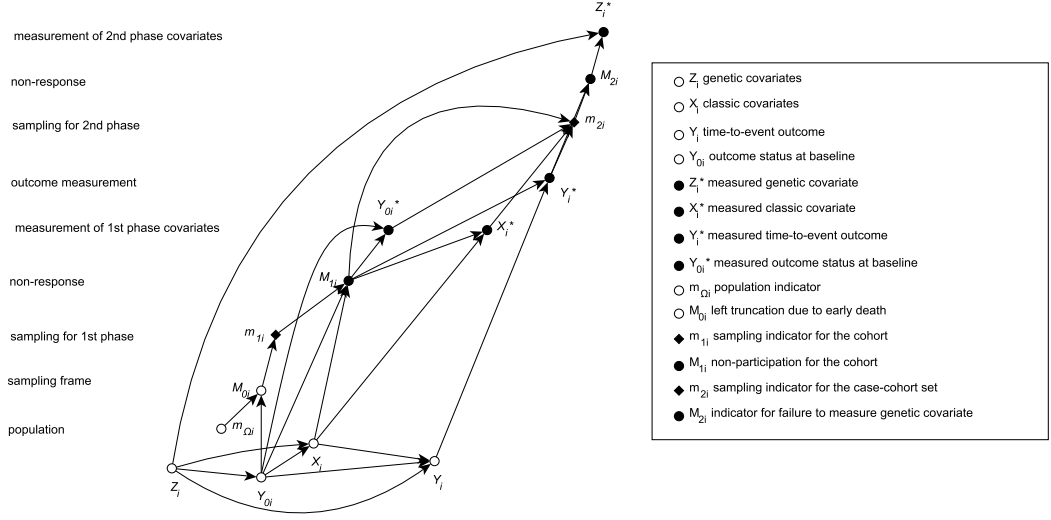}
\end{center}
\caption{Causal model with design for the two-stage case-cohort design used in the MORGAM Project \citep{morgamcasecohort,evans:2005}.  The sampling frame $\{i:M_{0i}=1\}$ is conditioned on the health status $Y_{0i}$ at the beginning of the study and this dependence must be taken into account when estimates for the population $\{i:m_{\Omega i}=1\}$ are required. At the first stage of the study, a random sample $\{i:m_{1i}=1\}$ is selected. The decision to participate $M_{1i}$ may depend on classic risk factors and current health status.  Classic risk factors $X_i^*$ and current health status $Y_{0i}^*$ are measured at the beginning of the study for the cohort members $\{i:M_{1i}=1\}$. Blood samples taken at the baseline are frozen to be used later. After a follow-up period of 10 years or more, the selection for the second stage is made on the basis of the measurements $X_i^*$ and $Y_{i}^*$. All disease cases and an age-stratified random subset of the cohort are selected to the case-cohort set $\{i:m_{2i}=1\}$ for which genetic factors $Z_{i}^*$ are measured. Nonresponse $M_{2i}$ occurs due to missing or contaminated samples or other technical reasons. \label{fig:morgamcasecohort}}
\end{figure}

\section{Discussion} \label{sec:discussion}
Causal models with design offer a systematic and unifying view to scientific inference. They present the causal assumptions, the study design and the data collection in a way that accounts for the complexity encountered in real-world problems. The examples in Section~\ref{sec:examples} demonstrate how the concept can be used to describe medical studies with multiple stages. Conclusions whether a causal or observational relationship can be estimated from the collected incomplete data can be made directly from the graph as it was demonstrated with the MORGAM Project. Despite the complex design, the estimation of the causal effects can be carried out in a systematic way via causal calculus as illustrated in Section~\ref{sec:estimation}.

Causal models with design present the population and the selection as intrinsic parts of the model. Selection nodes may have both incoming and outgoing connections to other nodes.  A distinction is made between a random variable and its measured value. Combined with the selection this allows the description of various sampling and missing data setups in terms of causal effects. 
 
The limitations of the causal model with design are in many ways similar to the limitations of the causal models in general. The presentation of causal assumptions in the form of a graphical model has the benefit that many problems can be solved without specifying the parameters of the model. On the other hand, the explicit parametric definition of the functional relationships is still the only decisive presentation of the model. Certain causal effects may be identifiable only under specific parametric assumptions such as linearity of the effect. 

The implications of the concept are two-fold. First, it ties together causality and study design and opens new possibilities for the practical application of graphical models. Second, it shows the key elements of the study in a compact visual format and thus increases the clarity and speed of communication. High standards of design, analysis and communication of scientific studies will significantly reduce the time and effort needed for the synthesis of scientific knowledge.

\section*{Acknowledgement}
The author thanks Olli Saarela, Mervi Eerola, Antti Penttinen, Jukka Nyblom, Jaakko Reinikainen and the anonymous referees for their comments that helped to improve the article. Kari Kuulasmaa is acknowledged for the numeric details in the description of the MORGAM Project.

\bibliographystyle{apalike}
\bibliography{dag,omat,surv}

\newcommand{\noopsort}[1]{} \newcommand{\printfirst}[2]{#1}
  \newcommand{\singleletter}[1]{#1} \newcommand{\switchargs}[2]{#2#1}
\begin{thebibliography}{}

\bibitem[Alkerwi et~al., 2010]{alkerwi2010comparison}
Alkerwi, A., Sauvageot, N., Couffignal, S., Albert, A., Lair, M.-L., and
  Guillaume, M. (2010).
\newblock Comparison of participants and non-participants to the {ORISCAV-LUX}
  population-based study on cardiovascular risk factors in {L}uxembourg.
\newblock {\em BMC Medical Research Methodology}, 10(1):80.

\bibitem[Angrist et~al., 1996]{angrist1996identification}
Angrist, J.~D., Imbens, G.~W., and Rubin, D.~B. (1996).
\newblock Identification of causal effects using instrumental variables (with
  comments).
\newblock {\em Journal of the American Statistical Association},
  91(434):444--472.

\bibitem[Asplund et~al., 2009]{stroke}
Asplund, K., Karvanen, J., Giampaoli, S., Jousilahti, P., Niemel{\"a}, M.,
  Broda, G., Cesana, G., Dallongeville, J., Ducimetriere, P., Evans, A., ,
  Ferrières, J., Haas, B., Jorgensen, T., Tamosiunas, A., D.Vanuzzo, Wiklund,
  P.-G., Yarnell, J., Kuulasmaa, K., and {Kulathinal, for the MORGAM Project},
  S. (2009).
\newblock Relative risks for stroke by age, sex, and population based on
  follow-up of 18 {E}uropean populations in the {MORGAM} {P}roject.
\newblock {\em Stroke}, 40(7):2319--2326.

\bibitem[Bareinboim and Pearl, 2012a]{Bareinboim:zidentifiability}
Bareinboim, E. and Pearl, J. (2012a).
\newblock Causal inference by surrogate experiments: z-identifiability.
\newblock In {de Freitas}, N. and Murphy, K., editors, {\em Proceedings of the
  Twenty-Eight Conference on Uncertainty in Artificial Intelligence}, pages
  113--120. AUAI Press.

\bibitem[Bareinboim and Pearl, 2012b]{Bareinboim:controllingselectionbias}
Bareinboim, E. and Pearl, J. (2012b).
\newblock Controlling selection bias in causal inference.
\newblock In {\em JMLR Proceedings of the Fifteenth International Conference on
  Artificial Intelligence and Statistics (AISTATS)}, volume~22, pages 100--108.

\bibitem[Bareinboim and Pearl, 2013a]{bareinboim2013general}
Bareinboim, E. and Pearl, J. (2013a).
\newblock A general algorithm for deciding transportability of experimental
  results.
\newblock {\em Journal of Causal Inference}, 1(1):107--134.

\bibitem[Bareinboim and Pearl, 2013b]{Bareinboim:metatransportability}
Bareinboim, E. and Pearl, J. (2013b).
\newblock Meta-transportability of causal effects: A formal approach.
\newblock In {\em Proceedings of the 16th International Conference on
  Artificial Intelligence and Statistics (AISTATS)}, pages 135--143.

\bibitem[Chou et~al., 1997]{chou1997characteristics}
Chou, P., Kuo, H.-S., Chen, C.-H., and Lin, H.-C. (1997).
\newblock Characteristics of non-participants and reasons for non-participation
  in a population survey in {Kin-Hu}, {Kinmen}.
\newblock {\em European Journal of Epidemiology}, 13(2):195--200.

\bibitem[Cohen and Duffy, 2002]{cohen2002nonrespondents}
Cohen, G. and Duffy, J.~C. (2002).
\newblock Are nonrespondents to health surveys less healthy than respondents?
\newblock {\em Journal of Official Statistics}, 18(1):13--24.

\bibitem[Cooper, 2000]{Cooper:bayesian}
Cooper, G.~F. (2000).
\newblock A {B}ayesian method for causal modeling and discovery under
  selection.
\newblock In Boutilier, C. and Goldszmidt, M., editors, {\em Proceedings of
  16th Conference on Uncertainty in Artificial Intelligence}, pages 98--106.

\bibitem[Didelez et~al., 2010]{Didelez:graphical}
Didelez, V., Kreiner, S., and Keiding, N. (2010).
\newblock Graphical models for inference under outcome-dependent sampling.
\newblock {\em Statistical Science}, 25(3):368--387.

\bibitem[Drivsholm et~al., 2006]{drivsholm2006representativeness}
Drivsholm, T., Eplov, L.~F., Davidsen, M., J{\o}rgensen, T., Ibsen, H.,
  Hollnagel, H., and Borch-Johnsen, K. (2006).
\newblock Representativeness in population-based studies: a detailed
  description of non-response in a {D}anish cohort study.
\newblock {\em Scandinavian Journal of Public Health}, 34(6):623--631.

\bibitem[Evans et~al., 2005]{evans:2005}
Evans, A., Salomaa, V., Kulathinal, S., Asplund, K., Cambien, F., Ferrario, M.,
  Perola, M., Peltonen, L., Shields, D., Tunstall-Pedoe, H., and {K. Kuulasmaa
  for The MORGAM Project} (2005).
\newblock {MORGAM} (an international pooling of cardiovascular cohorts).
\newblock {\em International Journal of Epidemiology}, 34:21--27.

\bibitem[Geneletti et~al., 2009]{Geneletti:adjusting}
Geneletti, S., Richardson, S., and Best, N. (2009).
\newblock Adjusting for selection bias in retrospective case-control studies.
\newblock {\em Biostatistics}, 10(1):17--31.

\bibitem[Hara et~al., 2002]{hara2002comparison}
Hara, M., Sasaki, S., Sobue, T., Yamamoto, S., and Tsugane, S. (2002).
\newblock Comparison of cause-specific mortality between respondents and
  nonrespondents in a population-based prospective study: ten-year follow-up of
  {JPHC Study Cohort I}.
\newblock {\em Journal of Clinical Epidemiology}, 55(2):150--156.

\bibitem[Heckman, 1979]{Heckman:sampleselection}
Heckman, J. (1979).
\newblock Sample selection bias as a specification error.
\newblock {\em Econometrica}, 47(1):153--161.

\bibitem[Huang and Valtorta, 2006]{Pearlsiscomplete}
Huang, Y. and Valtorta, M. (2006).
\newblock Pearl's calculus of intervention is complete.
\newblock In {\em Proceedings of the Twenty-Second Conference on Uncertainty in
  Artificial Intelligence}, pages 217--224. AUAI Press.

\bibitem[Jousilahti et~al., 2005]{jousilahti2005total}
Jousilahti, P., Salomaa, V., Kuulasmaa, K., Niemel{\"a}, M., and Vartiainen, E.
  (2005).
\newblock Total and cause specific mortality among participants and
  non-participants of population based health surveys: a comprehensive follow
  up of 54 372 {F}innish men and women.
\newblock {\em Journal of Epidemiology and Community Health}, 59(4):310--315.

\bibitem[Karvanen et~al., 2009a]{twostage}
Karvanen, J., Kulathinal, S., and Gasbarra, D. (2009a).
\newblock Optimal designs to select individuals for genotyping conditional on
  observed binary or survival outcomes and non-genetic covariates.
\newblock {\em Computational Statistics \& Data Analysis}, 53:1782--1793.

\bibitem[Karvanen et~al., 2010]{npmi}
Karvanen, J., Saarela, O., and Kuulasmaa, K. (2010).
\newblock Nonparametric multiple imputation of left censored event times in
  analysis of follow-up data.
\newblock {\em Journal of Data Science}, 8:151--172.

\bibitem[Karvanen et~al., 2009b]{newlyidentified}
Karvanen, J., Silander, K., Kee, F., Tiret, L., Salomaa, V., Kuulasmaa, K.,
  Wiklund, P.-G., Virtamo, J., Saarela, O., Perret, C., Perola, M., Peltonen,
  L., Cambien, F., Erdmann, J., Samani, N.~J., Schunkert, H., and {Evans for
  the MORGAM Project}, A. (2009b).
\newblock The impact of newly-identified loci on coronary heart disease, stroke
  and total mortality in the {MORGAM} prospective cohorts.
\newblock {\em Genetic Epidemiology}, 33:237--246.

\bibitem[Knol et~al., 2008]{Knol:whatdocasecontrol}
Knol, M.~J., Vandenbroucke, J.~P., Scott, P., and Egger, M. (2008).
\newblock What do case-control studies estimate? survey of methods and
  assumptions in published case-control research.
\newblock {\em American Journal Epidemiology}, 168(9):1073--1081.

\bibitem[Knudsen et~al., 2010]{knudsen2010health}
Knudsen, A.~K., Hotopf, M., Skogen, J.~C., {\O}verland, S., and Mykletun, A.
  (2010).
\newblock The health status of nonparticipants in a population-based health
  study the {H}ordaland {H}ealth {S}tudy.
\newblock {\em American Journal of Epidemiology}, 172(11):1306--1314.

\bibitem[Kulathinal and Arjas, 2006]{kulathinal:2006}
Kulathinal, S. and Arjas, E. (2006).
\newblock Bayesian inference from case-cohort data with multiple end-points.
\newblock {\em Scandinavian Journal of Statistics}, 33:25--36.

\bibitem[Kulathinal et~al., 2007]{morgamcasecohort}
Kulathinal, S., Karvanen, J., Saarela, O., Kuulasmaa, K., and {for the MORGAM
  Project} (2007).
\newblock Case-cohort design in practice -- experiences from the {MORGAM
  Project}.
\newblock {\em Epidemiological Perspectives \& Innovations}, 4(1):15.

\bibitem[Langholz, 2007]{Langholz:useofcohort}
Langholz, B. (2007).
\newblock Use of cohort information in the design and analysis of case-control
  studies.
\newblock {\em Scandinavian Journal of Statistics}, 34:120--136.

\bibitem[Lauritzen et~al., 1990]{Lauritzen:independence}
Lauritzen, S., Dawid, A., Wen, B., and Leimer, H.-G. (1990).
\newblock Independence properties of directed {M}arkov fields.
\newblock {\em Networks}, 20:491--505.

\bibitem[Little and Rubin, 2002]{littlerubin2002}
Little, R. J.~A. and Rubin, D.~B. (2002).
\newblock {\em Statistical analysis with missing data}.
\newblock Wiley.

\bibitem[McNamee, 2002]{McNamee:optimaltwostage}
McNamee, R. (2002).
\newblock Optimal designs of two-stage studies for estimation of sensitivity,
  specificity and positive predictive value.
\newblock {\em Statistics in Medicine}, 21:3609--3625.

\bibitem[Miettinen, 2011]{Miettinen:termsandconcepts}
Miettinen, O.~S. (2011).
\newblock {\em Epidemiological research: terms and concepts}.
\newblock Springer, Dordrecht.

\bibitem[Mohan et~al., 2013]{Mohan:missingdata}
Mohan, K., Pearl, J., and Tian, J. (2013).
\newblock Graphical models for inference with missing data.
\newblock In {\em {Proceedings of Neural Information Processing Systems
  Conference (NIPS-2013)}}.

\bibitem[Moher et~al., 2010]{CONSORT_exp}
Moher, D., Hopewell, S., Schulz, K.~F., Montori, V., Gotzsche, P.~C.,
  Devereaux, P., Elbourne, D., Egger, M., and Altman, D.~G. (2010).
\newblock {CONSORT} 2010 explanation and elaboration: updated guidelines for
  reporting parallel group randomised trials.
\newblock {\em Journal of Clinical Epidemiology}, 63(8):e1--e37.

\bibitem[Pearl, 1995]{Pearl:1995a}
Pearl, J. (1995).
\newblock Causal diagrams for empirical research.
\newblock {\em Biometrika}, 82(4):669--710.

\bibitem[Pearl, 2009]{Pearl:book}
Pearl, J. (2009).
\newblock {\em Causality: Models, Reasoning, and Inference}.
\newblock Cambridge University Press, second edition.

\bibitem[Reilly, 1996]{Reilly:optimalsampling}
Reilly, M. (1996).
\newblock Optimal sampling strategies for two-stage studies.
\newblock {\em American Journal of Epidemiology}, 143(1):92--100.

\bibitem[Rosenbaum and Rubin, 1983]{Rosenbaum:propensityscore}
Rosenbaum, P.~R. and Rubin, D.~B. (1983).
\newblock The central role of the propensity score in observational studies for
  causal effects.
\newblock {\em Biometrika}, 70(1):41--55.

\bibitem[Rubin, 1976]{Rubin:inferenceandmissingdata}
Rubin, D.~B. (1976).
\newblock Inference and missing data.
\newblock {\em Biometrika}, 63(3):581--592.

\bibitem[Rubin, 2008]{Rubin:designtrumps}
Rubin, D.~B. (2008).
\newblock For objective causal inference, design trumps analysis.
\newblock {\em The Annals of Applied Statistics}, 2(3):808–--840.

\bibitem[Saarela et~al., 2009]{prevalenceincidence}
Saarela, O., Kulathinal, S., and Karvanen, J. (2009).
\newblock Joint analysis of prevalence and incidence data using conditional
  likelihood.
\newblock {\em Biostatistics}, 10:575--587.

\bibitem[Saarela et~al., 2012]{secondary}
Saarela, O., Kulathinal, S., and Karvanen, J. (2012).
\newblock Secondary analysis under cohort sampling designs using conditional
  likelihood.
\newblock {\em Journal of Probability and Statistics}, Article ID 931416:37
  pages.

\bibitem[Schulz et~al., 2010]{CONSORT}
Schulz, K.~F., Altman, D.~G., Moher, D., and {CONSORT Group} (2010).
\newblock {CONSORT 2010 Statement: U}pdated guidelines for reporting parallel
  group randomized trials.
\newblock {\em Annals of Internal Medicine}, 152(11):726--732.

\bibitem[Seaman et~al., 2013]{Seaman:MAR}
Seaman, S., Galati, J., Jackson, D., and Carlin, J. (2013).
\newblock What is meant by ``missing at random''?
\newblock {\em Statistical Science}, 28(2):257--268.

\bibitem[Shpitser and Pearl, 2006a]{shpitser2006identificationconditional}
Shpitser, I. and Pearl, J. (2006a).
\newblock Identification of conditional interventional distributions.
\newblock In {\em Proceedings of the Twenty-Second Conference on Uncertainty in
  Artificial Intelligence (UAI2006)}, pages 437--444. AUAI Press.

\bibitem[Shpitser and Pearl, 2006b]{shpitser2006identificationjoint}
Shpitser, I. and Pearl, J. (2006b).
\newblock Identification of joint interventional distributions in recursive
  semi-{M}arkovian causal models.
\newblock In {\em Proceedings of the Twenty-First National Conference on
  Artificial Intelligence}, pages 1219--1226. AAAI Press.

\bibitem[Shpitser and Pearl, 2007]{Shpitser:counterfactuals}
Shpitser, I. and Pearl, J. (2007).
\newblock What counterfactuals can be tested.
\newblock In {\em {Proceedings of Twenty Third Conference on Uncertainty in
  Artificial Intelligence}}, pages 352--359, Vancouver, Canada.

\bibitem[Tian and Pearl, 2002]{tian2002general}
Tian, J. and Pearl, J. (2002).
\newblock A general identification condition for causal effects.
\newblock In {\em Proceedings of the Eighteenth National Conference on
  Artificial Intelligence}, pages 567--573. AAAI Press/The MIT Press.

\bibitem[{Van~Gestel} et~al., 2000]{VanGestel:powerofselectivegenotyping}
{Van~Gestel}, S., Houwing-Duistermaat, J.~J., Adolfsson, R., van Duijn, C.~M.,
  and Broeckhoven, C.~V. (2000).
\newblock Power of selective genotyping in genetic association analyses of
  quantitative traits.
\newblock {\em Behaviour Genetics}, 30(2):141--146.

\bibitem[Vandenbroucke, 1991]{Vandenbroucke:retroprospect}
Vandenbroucke, J.~P. (1991).
\newblock Prospective or retrospective: what's in the name?
\newblock {\em British Medical Journal}, 302:249--250.

\bibitem[Vandenbroucke et~al., 2007]{STROBE_exp}
Vandenbroucke, J.~P., {von Elm}, E., Altman, D.~G., G{\o}tzsche, P.~C., Mulrow,
  C.~D., Pocock, S.~J., Poole, C., Schlesselman, J.~J., Egger, M., and {for the
  STROBE Initiative} (2007).
\newblock Strengthening the reporting of observational studies in epidemiology
  ({STROBE}): {E}xplanation and elaboration.
\newblock {\em Epidemiology}, 18(6):805--835.

\bibitem[{von Elm} et~al., 2007]{STROBE}
{von Elm}, E., Altman, D.~G., Egger, M., Pocock, S., G{\o}tzsche, P.,
  Vandenbroucke, J., and {for the STROBE Initiative} (2007).
\newblock The strengthening the reporting of observational studies in
  epidemiology ({STROBE}) statement: guidelines for reporting observational
  studies.
\newblock {\em Epidemiology}, 18(6):800--804.

\end{thebibliography}

\section*{Appendix: Likelihood factorizations}
In this section, likelihood functions are presented for the examples of Section~\ref{sec:examples}. The likelihood functions are derived for the population $\{i:m_{\Omega i}=1 \}$ with the size $N$ starting from the factorization that follows directly from the DAG. At the first step, the likelihood function is written assuming that all variables are observed for the whole population. The measurements are redundant in this case because they are deterministic functions of the causal variables and the selection variables. The measurements becomes explicit when the likelihood function is further factorized according to the selection variables. Finally, the likelihood of the observed data is obtained as an integral over the unknown causal variables. 

Parameters $\boldsymbol{\theta}$ define the distribution of the causal variables and parameters $\boldsymbol{\psi}$ define the distribution of the selection variables. A vectorized notation similar to $\mathbf{X}=(X_1,X_2,\ldots,X_N)^T$ is used for all variables and the distributions are defined with respect to the first argument unless otherwise specified.

%

The likelihood function for the MORGAM Project case-cohort design presented in Figure~\ref{fig:morgamcasecohort} has the form  
\begin{align*} 
& p(\mathbf{m}_{\Omega},\mathbf{M}_0,\mathbf{m}_1,\mathbf{M}_1,\mathbf{m}_2,\mathbf{M}_2,\mathbf{Z},\mathbf{X},\mathbf{Y}_0,\mathbf{Y},\mathbf{Z}^*,\mathbf{X}^*,\mathbf{Y}_0^*,\mathbf{Y}^* \mid \boldsymbol{\theta},\boldsymbol{\psi} )= \\ 
& \prod_{i=1}^{N} p(m_{\Omega i}) p(Z_i\mid \boldsymbol{\theta}) p(Y_{0i} \mid Z_i,\boldsymbol{\theta}) p(M_{0i} \mid m_{\Omega i},Y_{0i},\boldsymbol{\psi}) p(m_{1i} \mid M_{0i},\boldsymbol{\psi}) p(M_{1i} \mid m_{1i},X_i,Y_{0i},\boldsymbol{\psi}) \\
& \times  p(X_i \mid Z_i,\boldsymbol{\theta}) p(Y_i \mid Z_i,Y_{0i},X_i,\boldsymbol{\theta}) p(m_{2i} \mid M_{1i},Y_{0i},Y_i,X_i,\boldsymbol{\psi}) p(M_{2i} \mid m_{2i},\boldsymbol{\psi}) =
\\
& \prod_{\{i:M_{2i}=1 \}} p_Z(Z_i^* \mid \boldsymbol{\theta})p_{Y_0}(Y_{0i}^* \mid Z_i=Z_i^*,\boldsymbol{\theta})
p(M_{0i}=1 \mid m_{\Omega i},Y_{0i}=Y_{0i}^*,\boldsymbol{\psi}) p(m_{1i}=1 \mid M_{0i}=1,\boldsymbol{\psi}) \\
& \times p(M_{1i}=1 \mid m_{1i}=1,X_i=X_i^*,Y_{0i}=Y_{0i}^*,\boldsymbol{\psi}) p_X(X_i^* \mid Z_i=Z_i^*,\boldsymbol{\theta}) \\
& \times p_Y(Y_i^* \mid Z_i=Z_i^*,Y_{0i}=Y_{0i}^*,X_i=X_i^*,\boldsymbol{\theta}) \\
& \times p(m_{2i}=1 \mid M_{1i}=1,Y_{0i}^*,Y_i^*,X_i^*,\boldsymbol{\psi}) p(M_{2i}=1 \mid m_{2i}=1,\boldsymbol{\psi})
\\
& \prod_{\{i:M_{2i}=0,m_{2i}=1 \}} p(Z_i \mid \boldsymbol{\theta})p_{Y_0}(Y_{0i}^* \mid Z_i,\boldsymbol{\theta})
p(M_{0i}=1 \mid m_{\Omega i},Y_{0i}=Y_{0i}^*,\boldsymbol{\psi})p(m_{1i}=1 \mid M_{0i}=1,\boldsymbol{\psi}) \\
& \times p(M_{1i}=1 \mid m_{1i}=1,X_i=X_i^*,Y_{0i}=Y_{0i}^*,\boldsymbol{\psi}) p_X(X_i^* \mid Z_i,\boldsymbol{\theta}) p_Y(Y_i^* \mid Z_i,Y_{0i}=Y_{0i}^*,X_i=X_i^*,\boldsymbol{\theta}) \\
& \times p(m_{2i}=1 \mid M_{1i}=1,Y_{0i}^*,Y_i^*,X_i^*,\boldsymbol{\psi}) p(M_{2i}=0 \mid m_{2i}=1,\boldsymbol{\psi})
\\
& \prod_{\{i:m_{2i}=0,M_{1i}=1 \}} p(Z_i \mid \boldsymbol{\theta})p_{Y_0}(Y_{0i}^* \mid Z_i,\boldsymbol{\theta})
p(M_{0i}=1 \mid m_{\Omega i},Y_{0i}=Y_{0i}^*,\boldsymbol{\psi}) p(m_{1i}=1 \mid M_{0i}=1,\boldsymbol{\psi}) \\
& \times p(M_{1i}=1 \mid m_{1i}=1,X_i=X_i^*,Y_{0i}=Y_{0i}^*,\boldsymbol{\psi}) p_X(X_i^* \mid Z_i,\boldsymbol{\theta}) p_Y(Y_i^* \mid Z_i,Y_{0i}=Y_{0i}^*,X_i=X_i^*,\boldsymbol{\theta}) \\
& \times p(m_{2i}=0 \mid M_{1i}=1,Y_{0i}^*,Y_i^*,X_i^*,\boldsymbol{\psi})
\\
& \prod_{\{i:M_{1i}=0,m_{1i}=1 \}} p(Z_i \mid \boldsymbol{\theta})p(Y_{0i} \mid Z_i,\boldsymbol{\theta})
p(M_{0i}=1 \mid m_{\Omega i},Y_{0i},\boldsymbol{\psi})p(m_{1i}=1 \mid M_{0i}=1,\boldsymbol{\psi}) \\
& \times p(M_{1i}=0 \mid m_{1i}=1,X_i,Y_{0i},\boldsymbol{\psi}) p(X_i \mid Z_i,\boldsymbol{\theta}) p(Y_i \mid Z_i,Y_{0i},X_i,\boldsymbol{\theta}) 
\\
& \prod_{\{i:m_{1i}=0,M_{0i}=1 \}} p(Z_i \mid \boldsymbol{\theta})p(Y_{0i} \mid Z_i,\boldsymbol{\theta})
p(M_{0i}=1 \mid m_{\Omega i},Y_{0i},\boldsymbol{\psi})p(m_{1i}=0 \mid M_{0i}=1,\boldsymbol{\psi})  \\
& \times p(X_i \mid Z_i,\boldsymbol{\theta}) p(Y_i \mid Z_i,Y_{0i},X_i,\boldsymbol{\theta}) 
\\
& \prod_{\{i:M_{0i}=0 \}} p(Z_i \mid \boldsymbol{\theta})p(Y_{0i} \mid Z_i,\boldsymbol{\theta})
p(M_{0i}=0 \mid m_{\Omega i},Y_{0i},\boldsymbol{\psi}) p(X_i \mid Z_i,\boldsymbol{\theta}) p(Y_i \mid Z_i,Y_{0i},X_i,\boldsymbol{\theta}). 
\end{align*}
The likelihood of the observed data is obtained as an integral over the unknown variables $\mathbf{Z}$, $\mathbf{X}$, $\mathbf{Y}_0$ and $\mathbf{Y}$: 
\begin{align*} 
& p(\mathbf{m}_{\Omega},\mathbf{M}_0,\mathbf{m}_1,\mathbf{M}_1,\mathbf{m}_2,\mathbf{M}_2,\mathbf{Z}^*,\mathbf{X}^*,\mathbf{Y}_0^*,\mathbf{Y}^* \mid \boldsymbol{\theta},\boldsymbol{\psi} )= 
\\
& \prod_{\{i:M_{2i}=1 \}} p_Z(Z_i^* \mid \boldsymbol{\theta})p_{Y_0}(Y_{0i}^* \mid Z_i=Z_i^*,\boldsymbol{\theta})
p(M_{0i}=1 \mid m_{\Omega i},Y_{0i}=Y_{0i}^*,\boldsymbol{\psi})p(m_{1i}=1 \mid M_{0i}=1,\boldsymbol{\psi}) \\
& \times p(M_{1i}=1 \mid m_{1i}=1,X_i=X_i^*,Y_{0i}=Y_{0i}^*,\boldsymbol{\psi}) p_X(X_i^* \mid Z_i=Z_i^*,\boldsymbol{\theta}) \\
& \times p_Y(Y_i^* \mid Z_i=Z_i^*,Y_{0i}=Y_{0i}^*,X_i=X_i^*,\boldsymbol{\theta}) \\
& \times p(m_{2i}=1 \mid M_{1i}=1,Y_{0i}^*,Y_i^*,X_i^*,\boldsymbol{\psi}) p(M_{2i}=1 \mid m_{2i}=1,\boldsymbol{\psi})
\\
& \prod_{\{i:M_{2i}=0,m_{2i}=1 \}}  \int p(Z_i \mid \boldsymbol{\theta})p_{Y_0}(Y_{0i}^* \mid Z_i,\boldsymbol{\theta}) p_X(X_i^* \mid Z_i,\boldsymbol{\theta}) p_Y(Y_i^* \mid Z_i,Y_{0i}=Y_{0i}^*,X_i=X_i^*,\boldsymbol{\theta}) \dd Z_i \\
& \times p(M_{0i}=1 \mid m_{\Omega i},Y_{0i}=Y_{0i}^*,\boldsymbol{\psi}) p(m_{1i}=1 \mid M_{0i}=1,\boldsymbol{\psi}) \\
& \times p(M_{1i}=1 \mid m_{1i}=1,X_i=X_i^*,Y_{0i}=Y_{0i}^*,\boldsymbol{\psi}) \\
& \times p(m_{2i}=1 \mid M_{1i}=1,Y_{0i}^*,Y_i^*,X_i^*,\boldsymbol{\psi}) p(M_{2i}=0 \mid m_{2i}=1,\boldsymbol{\psi})
\\
& \prod_{\{i:M_{1i}=0,m_{1i}=1 \}} \int\! \int\! \int\! \int  p(Z_i \mid \boldsymbol{\theta})p(Y_{0i} \mid Z_i,\boldsymbol{\theta})
p(M_{0i}=1 \mid m_{\Omega i},Y_{0i},\boldsymbol{\psi})  p(X_i \mid Z_i,\boldsymbol{\theta}) \\
&\times p(Y_i \mid Z_i,Y_{0i},X_i,\boldsymbol{\theta}) p(M_{1i}=0 \mid m_{1i}=1,X_i,Y_{0i},\boldsymbol{\psi}) \,\dd Z_i\, \dd X_i\, \dd Y_{0i}\, \dd Y_i\, p(m_{1i}=1 \mid M_{0i}=1,\boldsymbol{\psi}) 
\\
& \prod_{\{i:m_{1i}=0,M_{0i}=1 \}}  \int\! \int\! \int\! \int p(Z_i \mid \boldsymbol{\theta})p(Y_{0i} \mid Z_i,\boldsymbol{\theta})
p(M_{0i}=1 \mid m_{\Omega i},Y_{0i},\boldsymbol{\psi}) p(X_i \mid Z_i,\boldsymbol{\theta}) \\
& \times p(Y_i \mid Z_i,Y_{0i},X_i,\boldsymbol{\theta}) \,\dd Z_i\, \dd X_i\, \dd Y_{0i}\, \dd Y_i\, p(m_{1i}=0 \mid M_{0i}=1,\boldsymbol{\psi})
\\
& \prod_{\{i:M_{0i}=0 \}} \int\! \int\! \int\! \int\! p(Z_i \mid \boldsymbol{\theta})p(Y_{0i} \mid Z_i,\boldsymbol{\theta})
p(M_{0i}=0 \mid m_{\Omega i},Y_{0i},\boldsymbol{\psi}) p(X_i \mid Z_i,\boldsymbol{\theta}) \\
& \times p(Y_i \mid Z_i,Y_{0i},X_i,\boldsymbol{\theta}) \,\dd Z_i\, \dd X_i\, \dd Y_{0i}\, \dd Y_i.
\end{align*}

The likelihood function for the clinical trial presented in Figure~\ref{fig:clinicaltrial} has the form
\begin{align*}
 &  p( \mathbf{m}_{\Omega},\mathbf{m}_1,\mathbf{m}_2,\mathbf{Z},\mathbf{X},\mathbf{Y},\mathbf{T},\mathbf{T}',\mathbf{Z}^*,\mathbf{X}^*,\mathbf{Y}^*,\mathbf{T}^* \mid \boldsymbol{\theta},\boldsymbol{\psi})= \\
 & \prod_{i=1}^{N}  p(m_{\Omega i}) p(m_{1i} \mid m_{\Omega i},\boldsymbol{\psi}) p(Z_i \mid \boldsymbol{\theta}) p(m_{2i} \mid m_{1i},Z_i,\boldsymbol{\psi}) p(X_i \mid \boldsymbol{\theta})  \\ 
 & \times p(T_i' \mid m_{2i},\boldsymbol{\psi})p(T_i \mid T_i',\boldsymbol{\theta}) p(Y_i \mid X_i,T_i,\boldsymbol{\theta}) =
 \\
 &  \prod_{\{i:m_{2i}=1 \}}  p(m_{1i}=1 \mid m_{\Omega i},\boldsymbol{\psi})p_Z(Z_i^* \mid \boldsymbol{\theta})
 p(m_{2i}=1 \mid m_{1i}=1,Z_i^*,\boldsymbol{\psi}) \\
 &  \times p_X(X_i^* \mid \boldsymbol{\theta})p(T_i' \mid m_{2i}=1,\boldsymbol{\psi})p_T(T_i^* \mid T_i',\boldsymbol{\theta})p_Y(Y_i^* \mid X_i=X_i^*,T_i=T_i^*,\boldsymbol{\theta}) 
 \\
  &  \prod_{\{i:m_{2i}=0,m_{1i}=1 \}}  p(m_{1i}=1 \mid m_{\Omega i},\boldsymbol{\psi})p_Z(Z_i^* \mid \boldsymbol{\theta})p(m_{2i}=0 \mid Z_i^*,\boldsymbol{\psi}) p_X(X_i \mid \boldsymbol{\theta}) \\
  & \times p_T(T_i \mid \boldsymbol{\theta}) p_Y(Y_i \mid X_i,T_i,\boldsymbol{\theta}) \\
    & \prod_{\{i:m_{1i}=0 \}}  p(m_{1i}=0 \mid m_{\Omega i},\boldsymbol{\psi})p_Z(Z_i \mid \boldsymbol{\theta})   p_X(X_i \mid \boldsymbol{\theta})p_T(T_i \mid \boldsymbol{\theta})p_Y(Y_i \mid X_i,T_i,\boldsymbol{\theta}).
\end{align*}
The likelihood of the observed data is obtained as an integral over the unknown variables $\mathbf{Z}$, $\mathbf{X}$, $\mathbf{T}$ and $\mathbf{Y}$: 
\begin{align*}
 & p( \mathbf{m}_{\Omega},\mathbf{m}_1,\mathbf{m}_2,\mathbf{T}',\mathbf{Z}^*,\mathbf{Y}^*,\mathbf{X}^*,\mathbf{T}^* \mid \boldsymbol{\theta},\boldsymbol{\psi})= \\
  & \prod_{\{i:m_{2i}=1 \}} p(m_{1i}=1 \mid m_{\Omega i},\boldsymbol{\psi})p_Z(Z_i^* \mid \boldsymbol{\theta})
 p(m_{2i}=1 \mid m_{1i}=1,Z_i^*,\boldsymbol{\psi}) p_X(X_i^* \mid \boldsymbol{\theta}) \\
 & \times p(T_i' \mid m_{2i}=1,\boldsymbol{\psi})p_T(T_i^* \mid T_i',\boldsymbol{\theta})p_Y(Y_i^* \mid X_i=X_i^*,T_i=T_i^*,\boldsymbol{\theta}) 
 \\
    & \prod_{\{i:m_{2i}=0,m_{1i}=1 \}} p(m_{1i}=1 \mid m_{\Omega i},\boldsymbol{\psi})p_Z(Z_i^* \mid \boldsymbol{\theta})p(m_{2i}=0 \mid Z_i^*,\boldsymbol{\psi})  
  \\
 & \prod_{\{i:m_{1i}=0 \}} p(m_{1i}=0 \mid m_{\Omega i},\boldsymbol{\psi}).
\end{align*}
Only the first part of the likelihood is needed to estimate the effect of the treatment to the outcome. 
 
The likelihood for the nested case-control study presented in Figure~\ref{fig:nestedcc} has the form 
\begin{align*} 
& p( \mathbf{m}_{\Omega},\mathbf{m}_1,\mathbf{m}_2,\mathbf{Z},\mathbf{X},\mathbf{Y},\mathbf{Z}^*,\mathbf{X}^*,\mathbf{Y}^* \mid \boldsymbol{\theta},\boldsymbol{\psi})= \\ 
& \prod_{i=1}^{N} p(m_{\Omega i}) p(m_{1i} \mid m_{\Omega i},\boldsymbol{\psi}) p(Z_i \mid \boldsymbol{\theta})
p(X_i \mid \boldsymbol{\theta}) p(Y_i \mid Z_i,X_i,\boldsymbol{\theta})\\
& \times  p(m_{2i} \mid m_{1i},X_i,Y_i,\mathbf{X},\mathbf{Y},\boldsymbol{\psi})p(M_{2i} \mid m_{2i},\boldsymbol{\psi})= 
\\
 & \prod_{\{i:M_{2i}=1 \}} p(m_{1i}=1 \mid m_{\Omega i},\boldsymbol{\psi}) p_Z(Z_i^* \mid \boldsymbol{\theta})
 p_X(X_i^* \mid \boldsymbol{\theta})p_Y(Y_i^* \mid Z_i=Z_i^*,X_i=X_i^*,\boldsymbol{\theta}) \\
 & \times p(m_{2i}=1 \mid m_{1i}=1,X_i^*,Y_i^*,\mathbf{X}^*,\mathbf{Y}^*,\boldsymbol{\psi})p(M_{2i}=1 \mid m_{2i}=1,\boldsymbol{\psi})
 \\
  & \prod_{\{i:M_{2i}=0,m_{2i}=1 \}} p(m_{1i}=1 \mid m_{\Omega i},\boldsymbol{\psi}) p(Z_i \mid \boldsymbol{\theta})
 p_X(X_i^* \mid \boldsymbol{\theta}) p_Y(Y_i^* \mid Z_i,X_i=X_i^*,\boldsymbol{\theta}) \\
 & \times p(m_{2i}=1 \mid m_{1i}=1,X_i^*,Y_i^*,\mathbf{X}^*,\mathbf{Y}^*,\boldsymbol{\psi}) p(M_{2i}=0 \mid m_{2i}=1,\boldsymbol{\psi})
 \\
 & \prod_{\{i:m_{2i}=0,m_{1i}=1 \}} p(m_{1i}=1 \mid m_{\Omega i},\boldsymbol{\psi}) p(Z_i \mid \boldsymbol{\theta})
 p_X(X_i^* \mid \boldsymbol{\theta})p_Y(Y_i^* \mid Z_i,X_i=X_i^*,\boldsymbol{\theta}) \\
 & \times p(m_{2i}=0 \mid m_{1i}=1,X_i^*,Y_i^*,\mathbf{X}^*,\mathbf{Y}^*,\boldsymbol{\psi})
 \\
 & \prod_{\{i:m_{1i}=0 \}} p(m_{1i}=0 \mid m_{\Omega i},\boldsymbol{\psi}) p(Z_i \mid \boldsymbol{\theta})
 p(X_i \mid \boldsymbol{\theta})p(Y_i \mid Z_i,X_i,\boldsymbol{\theta}).
\end{align*} 
The likelihood of the observed data is obtained as an integral over the unknown variables $\mathbf{Z}$, $\mathbf{X}$ and $\mathbf{Y}$  
\begin{align*} 
& p( \mathbf{m}_{\Omega},\mathbf{m}_1,\mathbf{m}_2,\mathbf{Z}^*,\mathbf{Y}^*,\mathbf{X}^* \mid \boldsymbol{\theta},\boldsymbol{\psi})= \\ 
 & \prod_{\{i:M_{2i}=1 \}} p(m_{1i}=1 \mid m_{\Omega i},\boldsymbol{\psi}) p_Z(Z_i^* \mid \boldsymbol{\theta})
 p_X(X_i^* \mid \boldsymbol{\theta})p_Y(Y_i^* \mid Z_i=Z_i^*,X_i=X_i^*,\boldsymbol{\theta}) \\
 & \times p(m_{2i}=1 \mid m_{1i}=1,X_i^*,Y_i^*,\mathbf{X}^*,\mathbf{Y}^*,\boldsymbol{\psi})p(M_{2i}=1 \mid m_{2i}=1,\boldsymbol{\psi})
 \\
  & \prod_{\{i:M_{2i}=0,m_{2i}=1 \}} p(m_{1i}=1 \mid m_{\Omega i},\boldsymbol{\psi})  \int p(Z_i \mid \boldsymbol{\theta})
 p_X(X_i^* \mid \boldsymbol{\theta})p_Y(Y_i \mid Z_i,X_i=X_i^*,\boldsymbol{\theta}) \dd Z_i \\
 & \times p(m_{2i}=1 \mid m_{1i}=1,X_i^*,Y_i^*,\mathbf{X}^*,\mathbf{Y}^*,\boldsymbol{\psi})p(M_{2i}=0 \mid m_{2i}=1,\boldsymbol{\psi})
 \\
 & \prod_{\{i:m_{2i}=0,m_{1i}=1 \}} p(m_{1i}=1 \mid m_{\Omega i},\boldsymbol{\psi}) \int p(Z_i \mid \boldsymbol{\theta})
 p_X(X_i^* \mid \boldsymbol{\theta})p_Y(Y_i \mid Z_i,X_i=X_i^*,\boldsymbol{\theta}) \dd Z_i \\
& \times p(m_{2i}=0 \mid m_{1i}=1,X_i^*,Y_i^*,\mathbf{X}^*,\mathbf{Y}^*,\boldsymbol{\psi})
 \\
 & \prod_{\{i:m_{1i}=0 \}} p(m_{1i}=0 \mid m_{\Omega i},\boldsymbol{\psi}).
\end{align*}  

\end{document}